\documentclass[journal]{IEEEtran}

\usepackage{graphicx}
\usepackage{amsmath}
\usepackage{amssymb}

\usepackage[caption=false,font=footnotesize]{subfig}

\begin{document}

\title{Microwave properties of superconducting \\ atomic-layer deposited TiN films}
\author{P.~C.~J.~J.~Coumou$^{\textrm{1,*}}$, M.~R.~Zuiddam$^{\textrm{1}}$, E.~F.~C.~Driessen$^{\textrm{1}}$, P.~J.~de Visser$^{\textrm{1,2}}$, J.~J.~A.~Baselmans$^{\textrm{2}}$, and T.~M.~Klapwijk$^{\textrm{1}}$
\thanks{$^{\textrm{1}}$Kavli Institute of Nanoscience, Delft University of Technology, Delft, The Netherlands}
\thanks{$^{\textrm{2}}$SRON National Institute for Space Research, Utrecht, The Netherlands}
\thanks{$^{\textrm{*}}$Corresponding author, e-mail: p.c.j.j.coumou@tudelft.nl}}

\markboth{}
{Shell \MakeLowercase{\textit{et al.}}: Bare Demo of IEEEtran.cls for Journals}
\maketitle

\begin{abstract}
We have grown superconducting TiN films by atomic layer deposition with thicknesses ranging from 6 to 89~nm. This deposition method allows us to tune the resistivity and critical temperature by controlling the film thickness. The microwave properties are measured, using a coplanar-waveguide resonator, and we find internal quality factors above a million, high sheet inductances (5.2-620~pH), and pulse response times up to 100~$\mu$s. The high normal state resistivity of the films ($>$~100~$\mu\Omega$cm) affects the superconducting state and thereby the electrodynamic response. The microwave response is modeled using a quasiparticle density of states modified with an effective pair-breaker, consistently describing the measured temperature dependence of the quality factor and the resonant frequency.
\end{abstract}

\begin{IEEEkeywords}
TiN, thin films, ALD, microwave resonator.
\end{IEEEkeywords}
%
%
\IEEEpeerreviewmaketitle

\section{Introduction}
\IEEEPARstart{T}{here} is increasing interest in using superconducting metal-nitride films. The high normal-state resistivity leads to intrinsic material properties that are attractive for applications. The high sheet inductance of TiN films is beneficial for lumped-element kinetic-inductance detectors, providing efficient far-infrared absorption and high sensitivity~\cite{hleduc2010}. Electronic localization effects in InO$_\mathrm{x}$, NbN, and TiN films are the main conditions for detecting coherent quantum phase slips~\cite{oastafiev2012}. Hybrid superconductor/spin systems profit from the excellent properties of coplanar-waveguide (CPW) resonators of NbTiN films withstanding in-plane magnetic fields up to 300~mT~\cite{vranjan2012}. Uniform high-quality films and control of the electronic parameters is essential for all these applications. 

Atomic-layer deposition (ALD) allows us to tune the resistivity and the critical temperature by controlling the film thickness. Sputtering sub-stoichiometric TiN films is an alternative approach for tuning the critical temperature~\cite{hleduc2010}, with uniformity still being a concern~\cite{mvissers2012}. Sputtering bilayers of Ti and TiN is a recent effort to circumvent this problem ~\cite{mvissers2012b}. Uniformity over a large surface is a natural advantage of ALD.
 
In this paper we study the microwave response of ALD TiN films with different thicknesses. The high normal-state resistivity of the films ($>$ 100~$\mu\Omega$cm) affects the superconducting state and thereby the electrodynamic response. We model the microwave response using a quasiparticle density of states modified with an effective pair breaker, consistently describing the measured temperature dependence of the quality factor and the resonant frequency. In addition, we studied relaxation times in the disordered TiN films, which is a relevant parameter for some applications.

\section{Film preparation and characterization}
\begin{center}
\begin{table*}[!t]
\renewcommand{\arraystretch}{1.1}
\caption{Deposition settings of a single ALD cycle (0.45~$\textrm{\AA}$/cycle)}
\label{table:1}
\centering
\begin{tabular}{c c c c c c c c c}
\hline\hline
Step & Name & Time & Pressure & ICP valve & Ar & TiCl$_4$ valve & N$_2$ & H$_2$\\
 & & (s) & (mTorr) & & (sccm) & & (sccm) & (sccm)\\
\hline
1 & Argon purge & 1.5 & 80 & closing & 150 & closed & 0 & 0\\
2 & TiCl$_4$ dose & 0.04 & 80 & closed & 150 & open & 0 & 0\\
3 & Argon purge & 2 & 80 & closed & 150 & closed & 0 & 0\\
4 & Settle nitridation gasses & 2 & 30 & open & 0 & closed & 40 & 4\\
5 & Ignite plasma (400~W) & 0.5 & ~50* & open & 0 & closed & 40 & 4\\
6 & Nitridation (400~W) & 6 & 15 & open & 0 & closed & 40 & 4\\
7 & Argon purge & 1.5 & 80 & open & 150 & closed & 0 & 0\\
\hline \hline
\multicolumn{9}{l}{* This pressure is not reached within 0.5 seconds.}
\end{tabular}
\end{table*}
\end{center}

\begin{figure}[!t]
\centering
\subfloat{\includegraphics[width=3.49in]{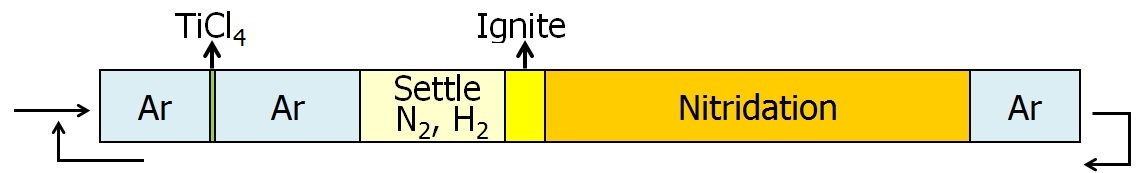}}\\
\subfloat{\includegraphics[width=2in]{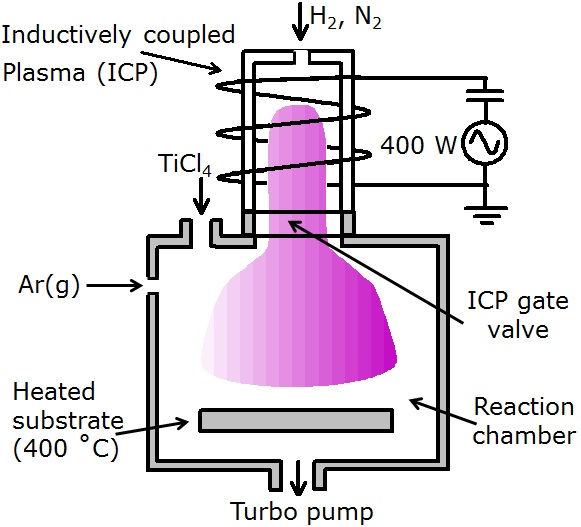}}
\subfloat{\raisebox{6mm}{\includegraphics[width=1.4in]{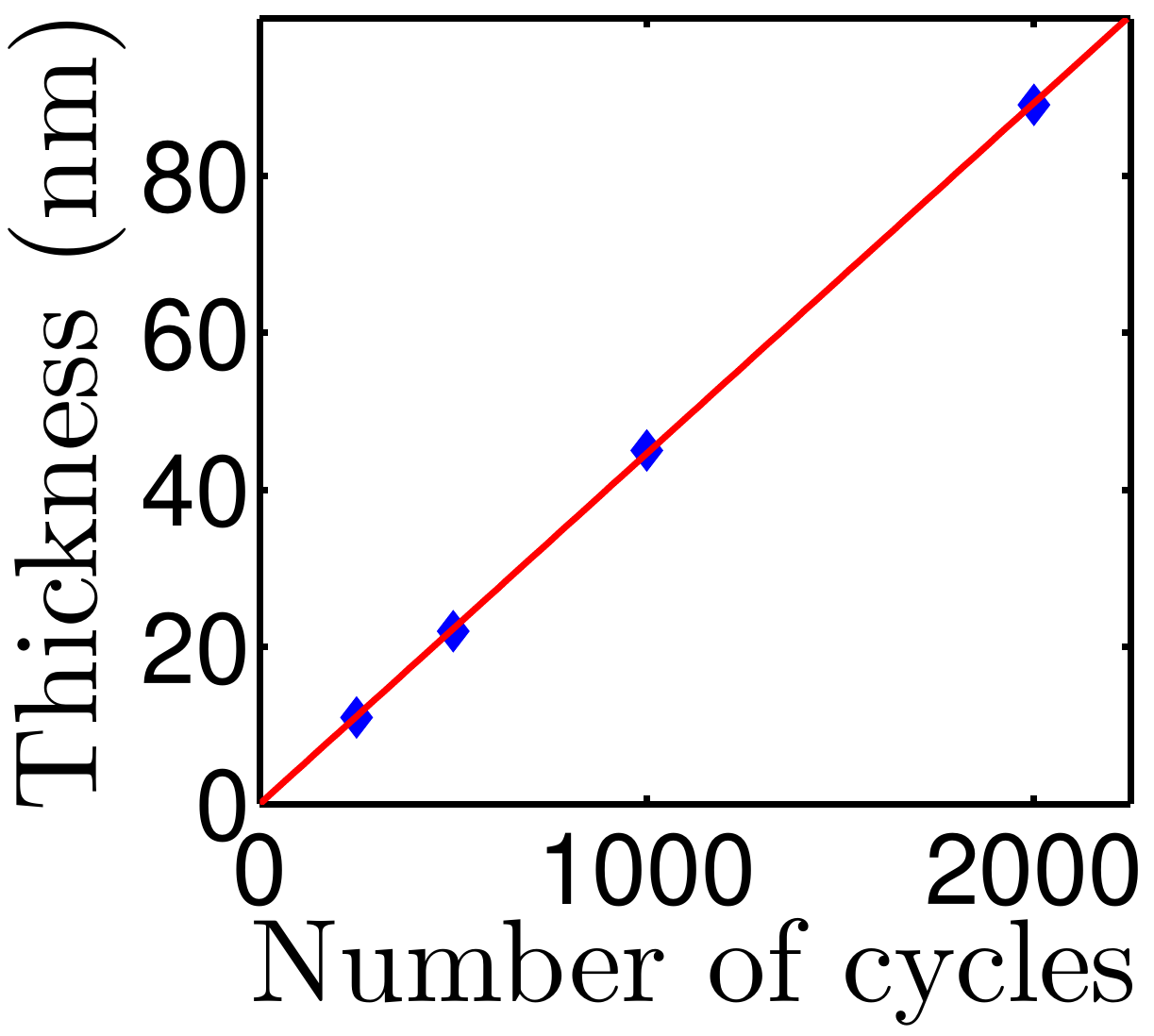}}}
\caption{(color online). A schematic representation of (left) the atomic-layer deposition (ALD) machine, and (top) the steps in a single ALD cycle with a total duration of 13.5 seconds. See the text and Table~\ref{table:1} for details on the process. Right: measured thickness as a function of number of deposition cycles, with a linear fit which has a slope of 0.45~$\textrm{\AA}$/cycle.}
\label{fig:1}
\end{figure}

Using plasma-assisted ALD, we grow superconducting TiN films. We make use of the FlexAL system of Oxford Instruments equipped with a (remote) inductively coupled plasma (ICP) source. A schematic representation of the machine is shown in Fig.~\ref{fig:1}. The ALD technique is based on sequential self-terminating gas-solid reactions at the surface of two gaseous precursors~\cite{rpuurunen2005}.

Process settings of a single ALD cycle are listed in Table~\ref{table:1}. A schematic representation of a single ALD cycle is shown in Fig.~\ref{fig:1}. Precursors TiCl$_4$ (step 2) and a plasma of H$_2$ and N$_2$ (step 4-6) are stepwise repetitively introduced into the reaction chamber and react into TiN and gaseous HCl. The used plasma power is 400 W. After each step, the residual gas is removed from the reaction chamber by an argon gas purge (step 3 and 7). The Ar, N$_2$, and H$_2$ gas flows are controlled with mass flow controllers and the pressure is regulated with a throttle valve. TiCl$_4$ is delivered by vapor draw, controlled with a fast open-close gas valve, which enables a short dose time of 40~ms. The Ar purge flow is not interrupted during this step (2). The ICP gate valve is closed (step 1) before the TiCl$_4$ dose in order to prevent the formation of a conductive layer of TiN in the ICP source. The substrate is heated to 400~$^\circ$C, while the walls of the reaction chamber are kept at 80~$^\circ$C.

Table~\ref{table:2} gives an overview of the films. The films are grown with a thickness $d$ ranging from 6 to 89~nm on high-resistivity ($\rho$ $>$ 10~k$\Omega$cm) Si (100) substrates with a thin native layer of silicon oxide on top. The thickness is measured with a step-profiler on a co-deposited lift-off structure. The value for the thinnest film is obtained by extrapolation. We obtain a linear growth rate of 0.45~$\textrm{\AA}$ per cycle (see Fig.~\ref{fig:1}), which is about one-tenth of the crystal lattice constant for TiN. The films are polycrystalline with a grain size $b$ increasing with increasing film thickness.

\begin{table}[!t]
\renewcommand{\arraystretch}{1.1}
\caption{Measured parameters of the films}
\label{table:2}
\centering
\begin{tabular}{c c c c c c c c}
\hline\hline
Film & $d$ & $b$ & $\rho$ & $k_Fl$ & $l$ & $T_c$ & $L_\mathrm{S}$\\
 & (nm) & (nm) &  ($\mu\Omega$cm) & & ($\textrm{\AA}$) & (K) & (pH)\\
\hline
A & 6* & 25 & 380 & 3.3 & 3.4 & 1.5 & 620\\
B & 11 & 27 & 356 & 3.5 & 3.5 & 2.2 & 200\\
C & 22 & 32 & 253 & 4.6 & 4.4 & 2.7 & 57\\
D & 45 & 37 & 187 & 6.1 & 5.7 & 3.2 & 18\\
E & 89 & 42 & 120 & 8.6 & 7.3 & 3.6 & 5.2\\
\hline \hline
\multicolumn{8}{l}{* Determined through extrapolation.}
\end{tabular}
\end{table}

\begin{figure}[!t]
\centering
\includegraphics[width=2.5in]{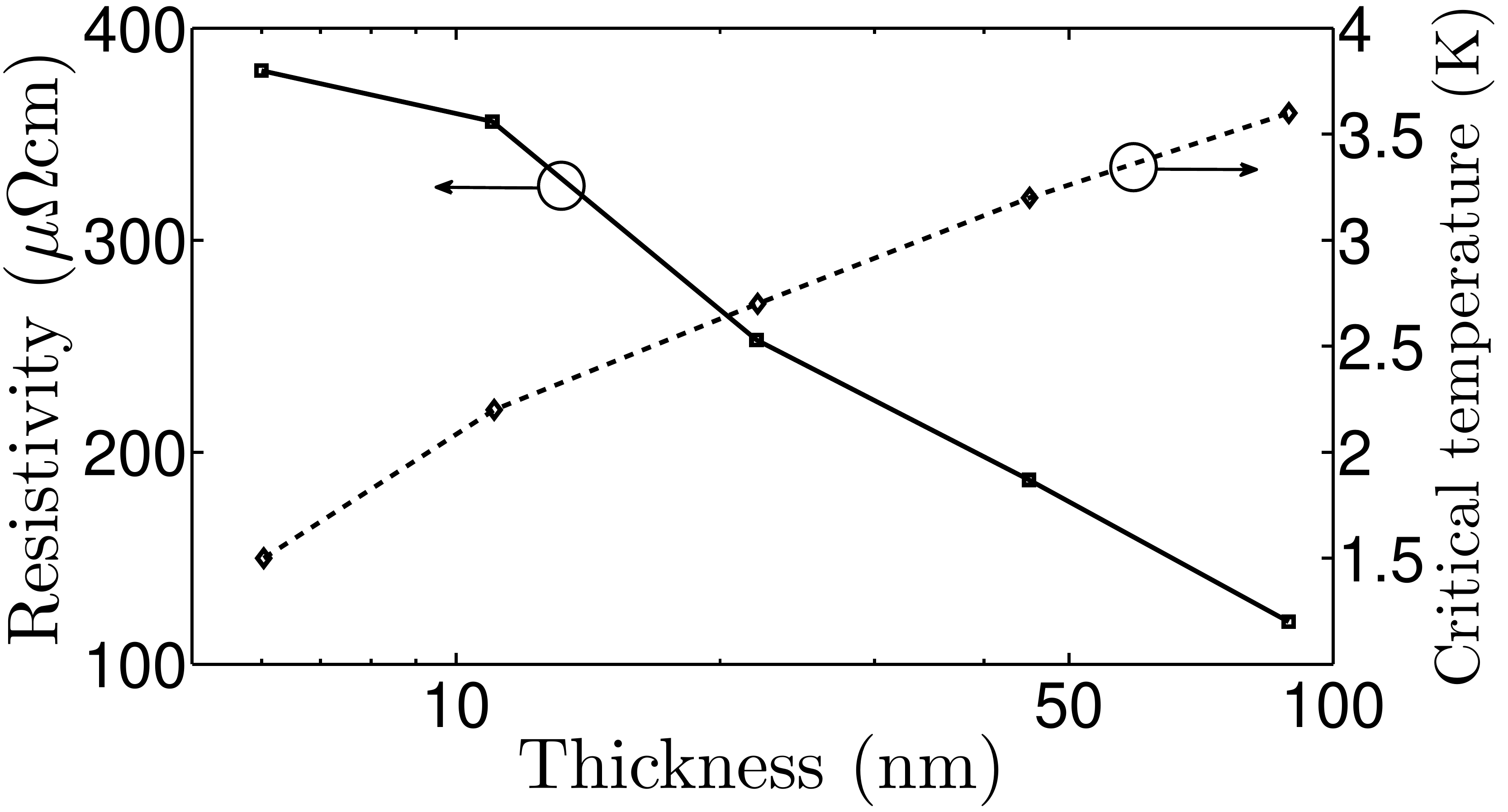}
\caption{Measured resistivity and critical temperature as a function of film thickness.}
\label{fig:2}
\end{figure}

Hall bar structures patterned in each film using e-beam lithography and reactive SF$_6$/O$_2$ ion etching are used to measure the critical temperature, and the sheet resistance $R_\Box$ and the Hall voltage $V_H(B)$ at 10~K. The resistivity $\rho=R_\Box d$ and the carrier density $n=IB/V_Hed$ follow directly, with $I$ and $B$ the applied current and magnetic field respectively. The Fermi wave vector and the elastic scattering length are respectively given by $k_F=\sqrt[3]{3\pi^2n}$ and $l=3\pi^2\hbar/e^2\rho(3\pi^2n)^{2/3}$ applying the free-electron theory~\cite{ashcroft:mermin}. The disorder increases, e.g. $k_Fl$ is decreasing, for reduced film thickness. 

We tune the normal-state and superconducting-state parameters, like the resistivity and the critical temperature by controlling the thickness. We show in Fig.~\ref{fig:2} that the resistivity increases and the critical temperature decreases with decreasing thickness. The reduction of critical temperature is understood from enhanced Coulomb interactions and a pair-breaking mechanism~\cite{edriessen2012}. 

\section{Microwave properties}
In order to measure the microwave properties of our films, we patterned CPW quarter-wave resonators in each film using e-beam lithography and reactive ion etching. The resonators are capacitively coupled to a CPW transmission line which is wire bonded to coaxial connectors. The sample is mounted onto the cold finger of a He-3 sorption cooler with a base temperature of 310~mK. A microwave signal from a vector network analyzer is fed to the sample through coaxial cables that are attenuated and filtered at 4~K. The amplified forward power transmission $S_{21}$ of the feed line is recorded as a function of temperature and microwave frequency.

Fig.~\ref{fig:3}~(inset~a) shows a typical resonance curve. The resonance line shape is asymmetric due to an impedance mismatch of the transmission line to the 50~$\mathrm{\Omega}$ coaxial cables. The resonant frequency $f_0$ and the internal quality factor $Q_\mathrm{i}$ are extracted from this line shape by fitting the relation~\cite{mkhalil2012}
\begin{equation}
\frac{1}{S_{21}} = 1+\frac{Q_\mathrm{i}/Q_\mathrm{c}}{1+2iQ_\mathrm{i}\frac{f-f_0}{f_0}},
\label{eq:S21}
\end{equation}
where $Q_\mathrm{c}$ is the complex coupling quality factor and $f$ is the microwave frequency. A small linear background transmission was accommodated in the fit to account for less-than-perfect calibration of the cables. The dashed curve is a fit of this line shape to the shown resonance.

We show a typical example of the response of the resonant frequency as function of temperature in Fig.~\ref{fig:3}. The measured resonant frequency is decreasing for increasing temperature, reflecting a change in the sheet inductance $L_\mathrm{S}$ of the superconductor. We describe the response to microwaves with the complex conductivity $\sigma = \sigma_1 - i\sigma_2$. The resonant frequency and the internal quality factor of a quarter-wave CPW resonator of a superconductor in the dirty limit follows from~\cite{rbarendsphd}
\begin{equation}
f_0 = \frac{1}{4 l \sqrt{[L_\mathrm{g}+\gamma L_\mathrm{S}(\sigma_2)]C}},
\label{eq:f0}
\end{equation}
and
\begin{equation}
Q_\mathrm{i} = \frac{2}{\chi\beta}\frac{\sigma_2}{\sigma_1},
\label{eq:qi}
\end{equation}
where $\beta=1+\frac{2d/\lambda}{\sinh(2d/\lambda)}$, with $\lambda=1/\sqrt{2\pi\mu_0f_0\sigma_2}$ the magnetic penetration depth and $d$ the film thickness, and where $\chi=\frac{L_\mathrm{S}}{L_\mathrm{S}+L_\mathrm{g}/\gamma}$ is the kinetic-inductance fraction, with $L_\mathrm{g}$ the geometric inductance of the CPW and $\gamma$ a geometric parameter converting sheet inductance to kinetic inductance~\cite{cholloway1995}. Further, $C$ is the geometric capacitance per unit length and $l$ the resonator length. The sheet inductance $L_\mathrm{S}$ is given~\cite{whenkels1977} by $\mu_0\lambda\coth(d/\lambda)$. 

Disorder is affecting the electrodynamics in a superconductor. The response of the resonant frequency as function of temperature deviates from Mattis-Bardeen theory~\cite{mattisbardeen} which assumes a textbook BCS density of states, as shown in Fig.~\ref{fig:3}. A consistent description is depicted by the solid line using a model discussed in Driessen {\textit{et al.}~\cite{edriessen2012}.

\begin{figure}[!t]
\centering
\includegraphics[width=3.3in]{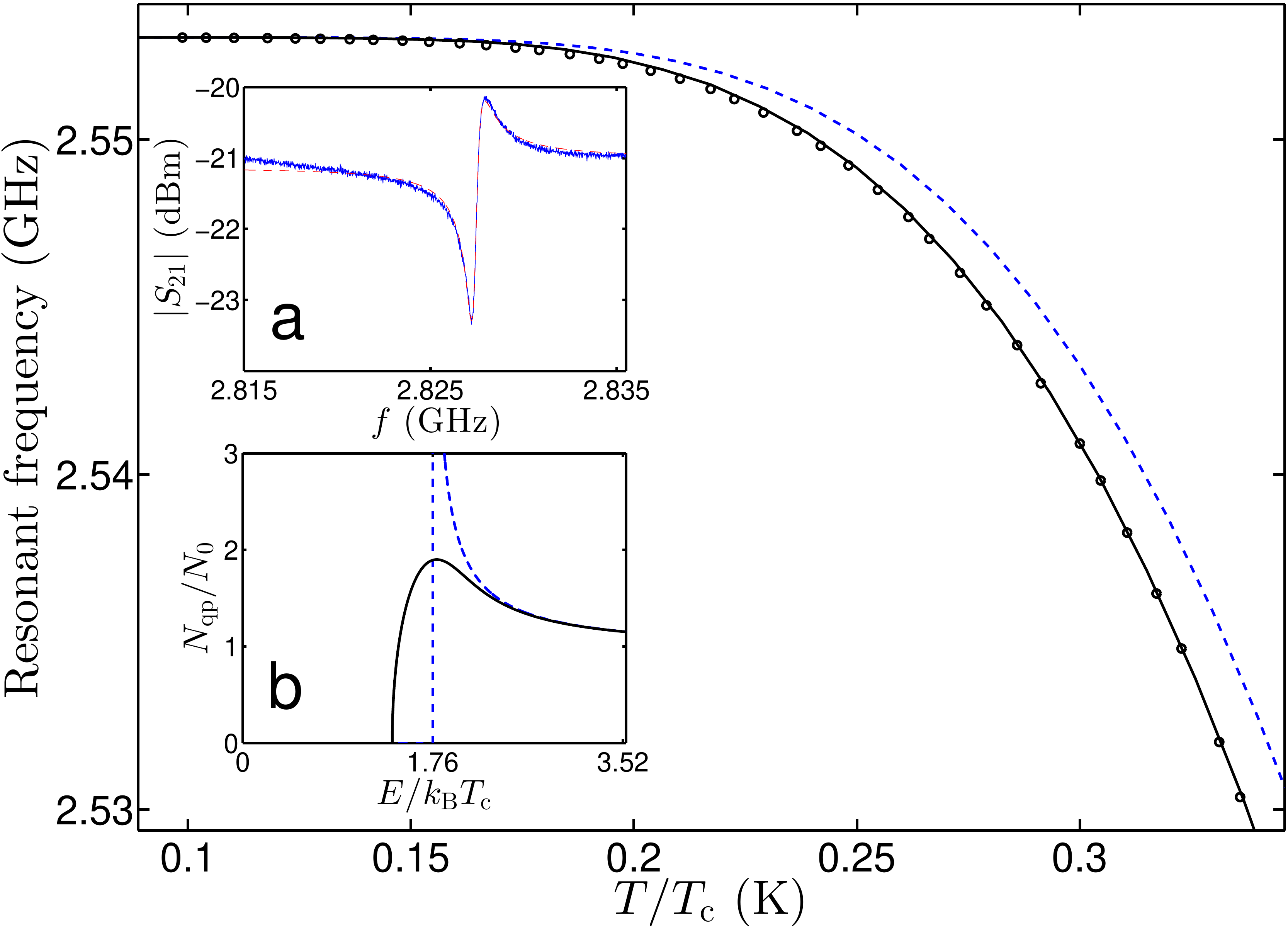}
\caption{(Color online) Typical example of the resonant frequency as a function of temperature (film D). Each point is determined from  a fit of (\ref{eq:S21}) to the measured resonance curves (inset~a). The dashed curve in the main figure is the calculated resonant frequency according to Mattis-Bardeen theory assuming a textbook BCS density of states. The solid curve is a fit using a broadened density of states (see~\cite{edriessen2012} for details). Inset (b) shows the differences in the quasiparticle densities of states used for calculating the resonant frequencies.}
\label{fig:3}
\end{figure}

The key element of the model is that we describe the  quasiparticle density of states with the Usadel equation modified with a pair-breaking mechanism, i.e. a spatially uniform breaking of time-reversal symmetry, the origin of which needs to be determined. Solutions to the Usadel equation provide the Greens functions $\sin\theta$ and $\cos\theta$ as function of energy, where $\theta$ is the pairing angle. The quasiparticle density of states follows from $N_\mathrm{qp} = N_0~\mathrm{Re}(\cos\theta)$, with $N_0$ the density of states in the normal metal at the Fermi energy. The coherence peak of the quasiparticle density of states is broadened when the pair-breaking is increased, which is visible in Fig.~\ref{fig:3}~(inset~b).

The complex conductivity can be calculated using a generalized Mattis-Bardeen equation~\cite{snam1967}, that is valid for a superconductor in the local limit with arbitrary pairing angle $\theta$:
\begin{eqnarray}
\hbar\omega\frac{\sigma_1}{\sigma_\mathrm{n}} = & \int_{-\hbar\omega}^{\infty}\mathrm{d}E~g_1(E,E')\left[1-2f(E')\right]\nonumber\\
&+2\int_0^\infty\mathrm{d}E~g_1(E,E')\left[f(E)-f(E')\right]
\label{eq:sigma1},
\end{eqnarray}
\begin{eqnarray}
\hbar\omega\frac{\sigma_2}{\sigma_\mathrm{n}} = & \int_{-\hbar\omega}^{\infty}\mathrm{d}E~g_2(E,E')\left[1-2f(E')\right]\nonumber\\
&+\int_0^\infty\mathrm{d}E~g_2(E',E)\left[1-2f(E)\right]
\label{eq:sigma2},
\end{eqnarray}
where $\sigma_\mathrm{n}$ is the normal state conductivity, $f(E)$ is the quasiparticle distribution function, and $E'=E+\hbar\omega$. The generalized coherence factors are given by $g_1(E,E')=\mathrm{Re}[\cos\theta(E)]\cdot\mathrm{Re}[\cos\theta(E')]+\mathrm{Re}[i\sin\theta(E)]\cdot\mathrm{Re}[i\sin\theta(E')]$, and $g_2(E,E')=\mathrm{Re}[\cos\theta(E)]\cdot\mathrm{Im}[\cos\theta(E')]+\mathrm{Im}[i\sin\theta(E)]\cdot\\\cdot\mathrm{Re}[i\sin\theta(E')]$. Without pair-breaking, (\ref{eq:sigma1}) and (\ref{eq:sigma2}) reduce to the standard Mattis-Bardeen expressions.

\begin{figure}[!t]
\centering
\includegraphics[width=3.3in]{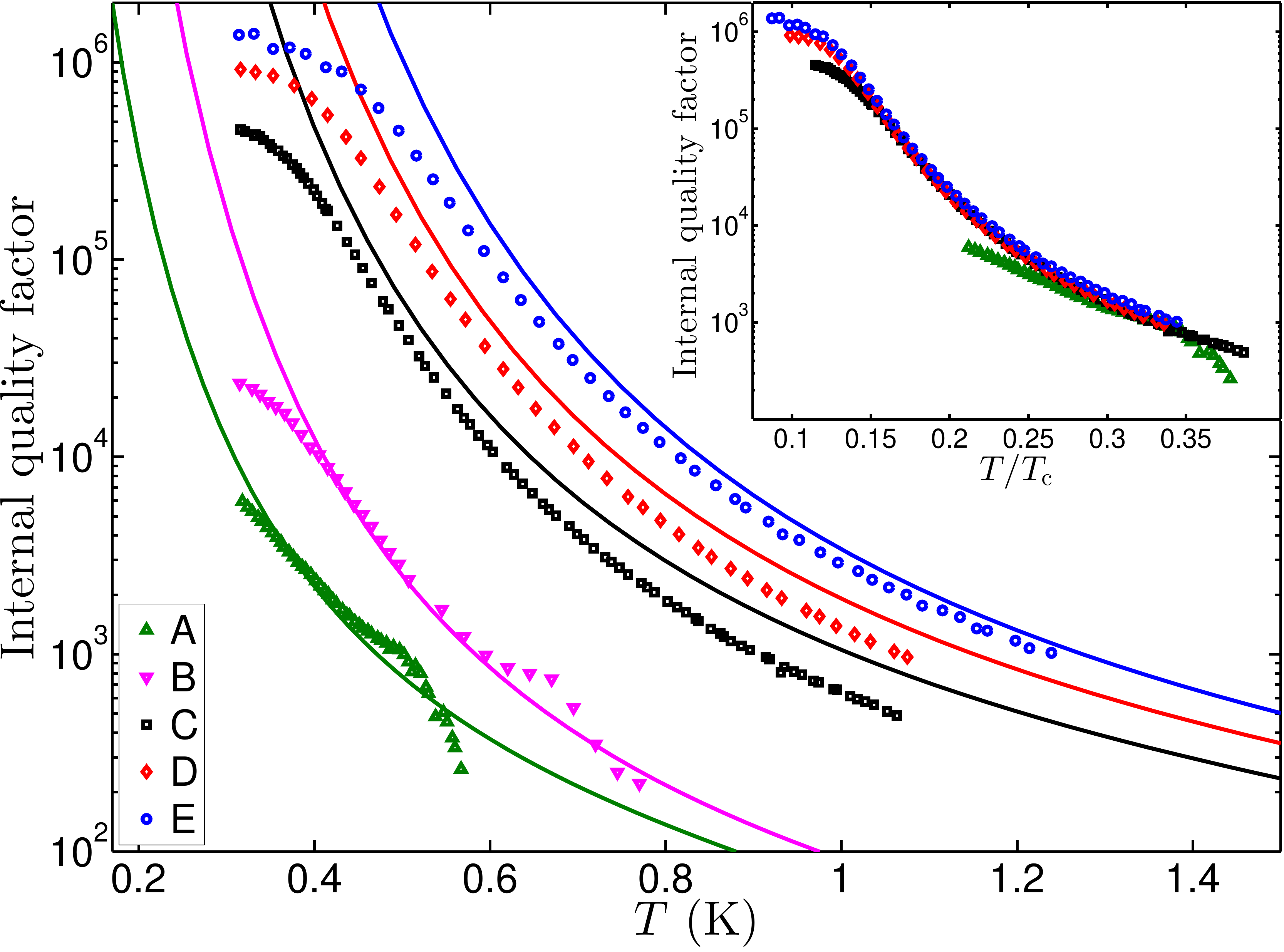}
\caption{(Color online) Measured internal quality factor as a function of temperature. The solid curves are calculated using the broadened density of states as determined from the fit to the resonant frequency response. There are no free parameters in these curves. Inset: measured internal quality factor as function of $T/T_\mathrm{c}$.}
\label{fig:4}
\end{figure}

The measured temperature dependence of the internal quality factor is displayed in Fig.~\ref{fig:4}. The solid curves are the predictions for the internal quality factors as function of temperature, using the broadened density of states as determined from the fit to the resonant frequency response. We focus on the temperature dependent part and exclude the saturation regime. The saturation regime is presently not understood and may partially be due to stray light and partially reflect a breakdown of the standard assumptions of the theory. The model is consistent with the measurements within a factor 2, without the use of any fitting parameter. The shape  of the curves only depends on $\sigma_1$. The temperature and disorder dependence of $\chi$, $\beta$, and $\sigma_2$ are negligible compared to $\sigma_1$.

We observe that the temperature dependence of the internal quality factor scales with the critical temperature (see inset Fig.~\ref{fig:4}). This implies that the quality factor can not be used as disorder sensitive probe for the superconducting state. The temperature dependence of $\sigma_1$ is dominated by the Fermi-Dirac function irrespective of the amount of disorder. Film B is not compared because the resonant frequency (9~GHz) is different from the other films which have resonant frequencies around 2-3~GHz. We observe that the saturation level lowers for decreasing film thickness. This might be understood from the fact that the thinner films have a smaller gap which reduces the required energy to create quasiparticle excitations.

\begin{figure}[!t]
\centering
\includegraphics[width=3.4in]{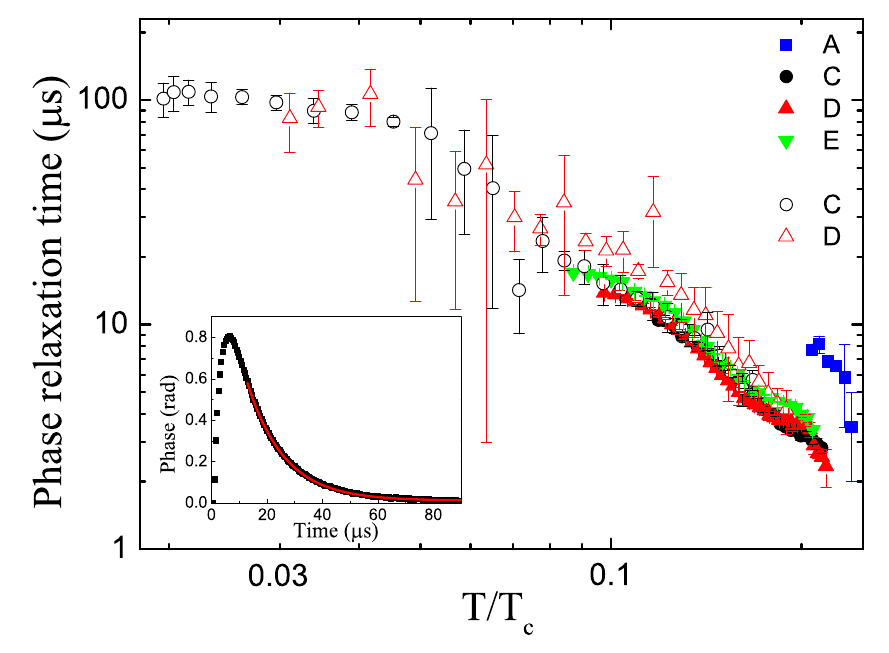}
\caption{(color online). Measured relaxation times of the resonator phase response as function of $T/T_\mathrm{c}$ for four of the films. Data with filled symbols are measured with an optical pulse in a He-3 cooler down to 310 mK. The inset shows a typical phase-response measurement with an exponential decay fit (13 us). Two of the films were measured with a microwave pulse down to 60 mK in a different setup (open symbols). The error bars represent statistical uncertainties obtained from the fitting procedure. }
\label{fig:5}
\end{figure}

We have measured relaxation times in the resonator by shining a pulse of visible light onto the sample and monitoring the resonator amplitude and phase as a function of time, while equilibrium is restored. For every temperature and power, we first measure the microwave response in a frequency sweep to obtain the resonance curve. We determine the resonant frequency and set the signal generator to that frequency. We monitor the resonator response at this frequency as a function of time after applying a 1~$\mathrm{\mu}$s pulse with an LED, with a maximum data acquisition rate of 2~MHz. This measurement is repeated 100 times, with a wait-time of 50~ms. The relaxation times are determined from a single exponential (decay) fit (inset Fig.~\ref{fig:5}) to the average of these 100 measurements in the regime of linear amplitude and phase response. Fig.~\ref{fig:5} shows the measured relaxation times of the phase response as a function of relative temperature for four of the films (closed symbols). To obtain data at lower temperatures, we have measured two of the films in an adiabatic demagnetization refrigerator down to 60 mK (open symbols in Fig.~\ref{fig:5}). This system is particularly stray-light tight, which is described in~\cite{jbaselmans2012}. Therefore, we can not use the light-pulse method and use alternatively a microwave-pulse technique. The larger uncertainty around $T/T_\mathrm{c}\approx0.06$ is due to lower response at these temperatures, which is observed for both samples.

We expect that the measured relaxation time is due to the recombination of the quasiparticles~\cite{rbarends2008c},~\cite{pdevisser2011}, since the applied pulse of energy breaks up Cooper-pairs into quasiparticles. The Kaplan framework~\cite{skaplan1976}, which is usually used to interpret recombination lifetimes, predicts that the leading temperature dependence of the lifetime is exponential. The measurements in Fig.~\ref{fig:5} show a much weaker temperature dependence, before saturating at 100 $\mu$s, and can therefore not be interpreted in this framework. On the other hand, we observe that the relaxation times for the different films scale similarly with $T/T_\mathrm{c}$, which suggests that the measured relaxation time is due to an intrinsic process in the film. As of yet, we do not have a method to calculate the quasiparticle lifetime including the broadened density of states.

\section{Conclusion}
Atomic layer deposition is an excellent way to grow superconducting TiN films with tunable resistivity and critical temperature by controlling the thickness. We measure internal quality factors above a million, high sheet inductances (5.2-620 pH) and pulse response times up to 100 $\mu$s using CPW microwave resonators. These values are in the same range as reported for sputtered TiN films~\cite{hleduc2010}, which makes ALD a good alternative for sputtering.

%
%
%
%
\section*{Acknowledgment}
The authors would like to thank R.~R.~Tromp for helpful discussions. The research was financially supported by the Dutch Foundation for Research of Matter (FOM) and MicroKelvin (Grant No.~228464, Capacities Specific Programme).
%
\ifCLASSOPTIONcaptionsoff
  \newpage
\fi
\end{document}